\documentclass[prl,showpacs,twocolumn,superscriptaddress,aps,a4paper]{revtex4-1}
\usepackage{dcolumn}
\usepackage{amsmath}
\usepackage{graphicx}
\usepackage{latexsym}
\usepackage{amsfonts}
\usepackage{amssymb}
\DeclareGraphicsExtensions{.pdf,.gif,.jpg}

\def\a{\alpha}
\def\b{\beta}

\def\l{\lambda}

\def\q{\psi}
\def\Q{\Psi}
\def\r{\rho}

\def\bra{\langle}
\def\ket{\rangle}

\def\ra{\rightarrow}
\def\inf{\infty}

\newcommand{\be}{\begin{equation}}
\newcommand{\ee}{\end{equation}}
\newcommand{\beq}{\begin{eqnarray}}
\newcommand{\eeq}{\end{eqnarray}}

\begin{document}

\title{Correlation induced memory effects in the  
transport properties of low dimensional systems}

\author{E. Perfetto}
\affiliation{Dipartimento di Fisica, Universit\'a di Roma Tor
Vergata, Via della Ricerca Scientifica 1, I-00133 Rome, Italy}

\author{G. Stefanucci}
\affiliation{Dipartimento di Fisica, Universit\'a di Roma Tor
Vergata, Via della Ricerca Scientifica 1, I-00133 Rome, Italy}
\affiliation{European Theoretical Spectroscopy Facility (ETSF)}

\author{M. Cini}
%\email{}
\affiliation{Dipartimento di Fisica, Universit\'a di Roma Tor
Vergata, Via della Ricerca Scientifica 1, I-00133 Rome, Italy}
\affiliation{Consorzio Nazionale Interuniversitario per le Scienze
Fisiche della Materia, Unit\'a Tor Vergata, Via della Ricerca
Scientifica 1, 00133 Rome, Italy}

\begin{abstract}
We demonstrate the remnant presence of {\em initial} correlations in
the {\em steady-state} electrical current flowing between 
low-dimensional interacting leads. The leads are
described as Luttinger liquids and electrons can tunnel via a quantum 
point-contact. We derive an analytic result for the time-dependent 
current and show that ground-state correlations have a large impact on 
the relaxation and long-time behavior. In particular, the I-V  
characteristic cannot be reproduced by quenching the interaction in time.
% In particular, the sudden quenching of 
% the interaction  {\em changes the I-V  characteristic}.
We further present a universal formula of the steady-state current $j_{S}$ 
for an arbitrary sequence of interaction quenches. It is established that 
$j_{S}$ is history dependent provided that the switching process is 
non-smooth.

\end{abstract}

\maketitle

Non-equilibrium phenomena in open nanoscale systems offer
a formidable challenge to modern science\cite{bp.book}. 
Controlling the electron dynamics of a molecular device is the 
ultimate goal of nanoelectronics and quantum 
computation\cite{books}; its microscopic description 
a problem at the forefront of statistical quantum 
physics\cite{souza.2007}.
% The difficulties arise from the high non-homogeneity of the system 
% (molecule), the possibility for electrons to scatter directly (in the 
% molecule) or indirectly (in the baths), and the presence of 
% external time-dependent (TD) fields which drive the system out of equilibrium.  
Resorting to approximate methods is inevitable to progress.

Standard many-body techniques consider an initial state with {\em no 
interaction} and with {\em no contact} between the system and the baths 
(leads from now on), and then switch them on in 
time\cite{ccns.1971-meir,jauho,pss.2010}.
In fact, it is plausible to believe that starting from the true
{\em interacting} and {\em contacted} state 
the long-time results would  not change.
To what extent, however, such belief is actually the truth? 
This question is of both practical and fundamental 
interest.
It has been shown by us \cite{cini-stef} and others \cite{cnz.2009} that for 
non-interacting electrons the initial contact plays no role at the 
steady-state \cite{s.2007} (theorem of equivalence). Allowing for interactions
in the system only (non-interacting leads) My\"oh\"anen {\em et al.} 
found that steady-state quantities are not sensitive to initial 
correlations either\cite{myo}. It is the purpose of this Letter to show that
interacting leads change dramatically the picture:
the switching process can indeed
have a large impact on the {\em relaxation} 
and the {\em steady-state} behavior.

\begin{figure}[tbp]
\includegraphics[height=2.5cm ]{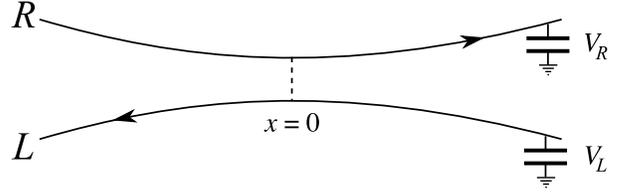}
\caption{Sketch of the device. Two interacting leads hosting $L$
and $R$ movers are connected at $x=0$ via a weak link. A
bias voltage $V_{L}-V_{R}$ can be applied between the leads.}
\label{device}
\end{figure}

We consider two one-dimensional 
interacting leads described as Luttinger Liquids 
(LL)\cite{giamarchi}, see Fig. \ref{device}.
It is known that a LL does not relax to the ground state after a 
sudden quench of the interaction \cite{cazalilla,perfetto,man-ekw.2009}
(thermalization breakdown). The implications of such important 
result in the context of time-dependent (TD) transport are totally 
unknown and will be here explored for the first time.
We compare the dynamics of initially (a) contacted ($\eta_{T}=1$) 
versus uncontacted ($\eta_{T}=0$) and (b) interacting 
($\eta_{I}=1$) versus non-interacting ($\eta_{I}=0$) 
LL when driven out of equilibrium by an external bias.
Our main findings are that in (a) the system relaxes towards the same steady-state
although with a {\em different power law} decay.  In (b) the sudden quench of 
the interaction when $\eta_{I}=0$ {\em alters the
steady-current $j_{S}$} as well. This remains true for an arbitrary 
sequence of interaction quenches. We are able to write $j_{S}$ as 
an explicit functional of the switching process and to establish that 
$j_{S}$ is history dependent for non-smooth switchings.

The equilibrium Hamiltonian for the system of Fig. \ref{device} reads 
\be
H_{0}=H_{R}+H_{L}+\eta_{I}H_{I}+\eta_{T}H_{T}. 
\label{h0}
\ee
The one-body part of the left ($L$) and right ($R$) 
leads is $H_{R/L}=\mp iv_{F}\int_{-\infty}^{\infty}dx \;
\psi^{\dagger}_{R/L}(x)\partial_{x} \psi_{R/L}(x)$,
where the fermion field $\psi_{R/L}$ describes right/left moving electrons 
in lead $R/L$ with Fermi velocity $v_{F}$ (chiral leads). We take a 
density-density interaction of the form 
$ H_{I}=\frac{1}{2}
\int_{-\infty}^{\infty}dx \, [2g_{2} \, \rho_{R}(x)
\rho_{L}(x)+g_{4}(\rho^{2}_{R}(x)+\rho^{2}_{L}(x))]$,
where $\rho_{R/L}\equiv\,\,:\psi^{\dagger}_{R/L} \psi_{R/L}:$ is (in standard 
notation) the
fermionic density operator relative to the Fermi sea, and
$g_{2/4}$ are the forward scattering couplings,
corresponding to inter/intra lead interactions respectively.
The two chiral liquids are linked at $x=0$ via the
tunneling term 
$H_{T}=\lambda
\psi^{\dagger}_{R}(0)\psi_{L}(0)+{\rm H.c.}$,
which does not commute with the
total number of electrons $N_{R/L}$ of each lead.

If a bias $V=V_{L}-V_{R}$ is applied at, say, 
time $t=0$, a
finite current $j(t)$ starts flowing across the link. The current 
operator (in atomic units) $J=dN_{L}/dt=-dN_{R}/dt$ reads
$J=i \lambda \psi^{\dagger}_{R}(0)\psi_{L}(0)+{\rm H.c.}$.
At zero temperature the current $j(t)$ is the TD average of $J$ over the ground state 
$|\Q_{0}\ket$ of $H_{0}$, i.e.,
\be 
j(t)= \bra\Q_{0}|J_{H_{1}}(t)|\Q_{0}\ket,
\label{ens}
\ee
where $J_{H_{1}}(t)$ is the $J$ operator in the Heisenberg
representation with respect to the interacting, contacted and biased 
Hamiltonian $H_{1}=H_{L}+H_{R}+H_{I}+H_{T}+H_{V}$, $H_{V}=V_{R}N_{R}+V_{L}N_{L}$. 
Note that the factors $\eta_{I},\eta_{T}$ refer to times $t<0$ 
and different values $\eta_{I},\eta_{T}=0,1$
yields different $H_{0}$ and hence different initial states 
$|\Q_{0}\ket$. At positive 
times the Hamiltonian is the same in all cases.

{\em The exact non-interacting solution.} We start our analysis by 
calculating $j(t)$ when $\eta_{T}=0$ (initially uncontacted) and $g_{2}=g_{4}=0$ 
(always non-interacting). In terms of the
Fourier transform $\psi_{k_{R/L}}$ of the original
fermion fields, the current operator reads
$J=(i\l/a)\sum_{kk'}\psi_{k_{R}}^{\dag}\q_{k'_{L}}+{\rm H.c.}$,
with $a$ the usual short-distance cutoff.
Its expectation value is then 
$j(t)=\lambda \mathrm{Im}  \sum_{\alpha}\int \frac{dp}{\pi} 
\Gamma_{p}^{R \alpha} (t)  f^{\a}_{p} \left[\Gamma_{p}^{L \alpha} (t) 
\right]^{*}$ where the sum runs over $\a=R,L$, 
$f^{R/L}_{p}=f(\pm v_{F} p)$ is the Fermi function of lead 
$R/L$ and $\Gamma_{p}^{\alpha \beta}(t)=-ia \int \frac{dk}{2\pi}
\bra\Q_{0}|\q_{k_{\a}}e^{-iH_{1}t}\q^{\dag}_{p_{\b}}|\Q_{0}\ket$
is the sum of the probability amplitudes (retarded Green's 
functions) for the transition $p_{\b}\ra k_{\a}$. 
From the Dyson equation it is straightforward to find  
$\Gamma^{\alpha \alpha}_{p}(t)=-i e^{i(\alpha v_{F} 
p +V_{\alpha})t}/(1+c^{2})$ and 
$\Gamma^{\bar{\alpha}\a}_{p}(t)= -ic \, 
\Gamma^{\a\a}_{p}(t)$, with $c=\lambda/(2v_{F})$, and hence
\be
j(t)=\frac{2c^{2}}{\pi(1+c^{2})^{2}} V  \, .
\label{curr0}
\ee 
The current is {\em discontinuous} in time; the steady-state value is 
reached instantaneously. This is due to the unbound 
(relativistic) energy spectrum\cite{jauho} and the lack of 
interactions, as discussed in detail in Ref. \onlinecite{swmk.2008}.
As we shall see, when $H_{I}\neq 0$ the transient regime is more 
complex. 

{\em Current to lowest order in $\l$.} The problem does not have 
an exact solution when both $H_{I}$ and $H_{T}$ are present. 
Below, we calculate $j(t)$ to lowest order in $\l$. In general, 
perturbative treatments in the tunneling amplitude are a delicate 
issue\cite{tunn1}. In our case $j(t)$ has a Taylor expansion 
with convergence radius $\l<2v_{F}$  for $H_{I}=0$, see Eq. (\ref{curr0}).
We, therefore, expect a finite convergence radius
at least for small interaction strengths. Let 
the unperturbed Hamiltonian be 
$\tilde{H}_{0}=H_{R}+H_{L}+\eta_{I}H_{I}$ 
in equilibrium ($t<0$) and 
$\tilde{H}_{1}=H_{R}+H_{L}+H_{I}+H_{V}$  
at positive times. At zero temperature and to lowest order in $\l$
\beq 
j(t) &=& i \langle \tilde{\Q}_{0}|\int_{0}^{t} ds \left[ 
H_{T ,\tilde{H}_{1}}(s),J_{\tilde{H}_{1}}(t) \right]
-\eta_{T}\int_{0}^{-i\infty}d\tau 
\nonumber \\
&\times& \left [ H_{T ,
\tilde{H}_{0}}(\tau) J_{\tilde{H}_{1}}(t) + J_{\tilde{H}_{1}}(t) H_{T 
,\tilde{H}_{0}}(-\tau) \right]
| \tilde{\Q}_{0} \rangle , 
\label{tdcurr}
\eeq
with $|\tilde{\Q}_{0}\ket$ the ground state of $\tilde{H}_{0}$.
The first term in the r.h.s. is the standard Kubo formula. 
Such term alone describes the transient response when the contacts
are switched on at $t=0$ ($\eta_{T}=0$). If, however, the equilibrium 
system is already contacted ($\eta_{T}=1$) we must account for 
a correction; this is the physical content of the second 
term\cite{note}. At any finite time
initial correlation effects are 
visible in both terms due to the ground state dependence on
$\eta_{I}$. 
When $t\ra\inf$ only the Kubo term survives, which 
yields the steady-current $j_{S}$. 
The dependence of $j_{S}$ on the ground state ($\eta_{I}=0,1$) will be 
addressed below.

% The averages in Eq. (\ref{tdcurr}) can be explicitly calculated by
% resorting to the bosonization method\cite{giamarchi}. 
% The Hamiltonian $H=H_{R}+H_{L}+H_{I}=\frac{v}{2} \int_{-\infty}^{\infty}dx [K^{-1}
% (\partial_{x}\phi(x))^{2}+K \partial_{x}\theta(x))^{2}]$  becomes a simple quadratic 
% form of the scalar fields $\phi$ and $\theta$ with commutation relation
% $[\phi(x),\theta(x')]=i \mathrm{sgn}(x-x')/2$. In the bosonized 
% Hamiltonian $v=\sqrt{(2\pi v_{F}+g_{4})^{2}-g^{2}_{2})}/2\pi$
% is the renormalized velocity and 
% $K=\sqrt{(2\pi v_{F}+g_{4}-g_{2})/(2\pi v_{F}+g_{4}+g_{2})}$ is a 
% parameter which measures the interaction strength. Note that
% $0<K\leq 1$ for repulsive interactions;  $K=1$ corresponds to the noninteracting case while
% small values of $K$ indicate a strongly correlated regime.
% The great advantage of the bosonization is that 
% the averages over $|\tilde{\Q}_{0}\ket$ 
% become averages over 
% the vacuum of the normal modes of $\phi$ and $\theta$.

The averages in Eq. (\ref{tdcurr}) can be explicitly calculated by
resorting to the bosonization method\cite{giamarchi}. 
We introduce the 
scalar fields $\phi$ and $\theta$ from 
$\rho_{R}(x)+\rho_{L}(x)=\frac{1}{\sqrt{\pi}}\partial_{x}\phi(x)$
and 
$\psi_{R/L}(x)=\frac{\kappa_{R/L}}{\sqrt{2\pi a} }
e^{i\sqrt{\pi}[\phi(x) \mp \theta(x)]}$, with $\kappa_{R/L}$ the 
anticommuting Klein factors. The scalar fields obey the commutation relation
$[\phi(x),\theta(x')]=i \mathrm{sgn}(x-x')/2$.
In terms of $\phi$ and $\theta$ the Hamiltonian $H=H_{R}+H_{L}+H_{I}$ is 
a simple quadratic form
$H=\frac{v}{2} \int_{-\infty}^{\infty}dx [K^{-1}
(\partial_{x}\phi(x))^{2}+K \partial_{x}\theta(x))^{2}]$, 
with $v=\sqrt{(2\pi v_{F}+g_{4})^{2}-g^{2}_{2})}/2\pi$
the renormalized velocity and 
$K=\sqrt{(2\pi v_{F}+g_{4}-g_{2})/(2\pi v_{F}+g_{4}+g_{2})}$ a 
parameter which measures the interaction strength. Note that
$0<K\leq 1$ for repulsive interactions;  $K=1$ corresponds to the noninteracting case while
small values of $K$ indicate a strongly correlated regime.

By employing the gauge transformation\cite{feldman} $\psi_{L,R} \rightarrow
\psi_{L,R}e^{iV_{L/R}t}$ the problem of evaluating Eq. (\ref{tdcurr})
is reduced to the calculation of different bosonic vacuum averages\cite{giamarchi}. After 
some tedious algebra one finds
\be
j(t)= \xi \,\mathrm{Re} \, [\eta_{T} A_{\eta_{I}}(t)+B_{\eta_{I}}(t)] \, ,
\label{curregen}
\ee
where
\beq
A_{0}(t)&=&\sin(Vt)\int_{0}^{\infty}d\tau \,\gamma^{2}(t+i\tau),
\nonumber \\
B_{0}(t) &=&i
\int_{0}^{t}ds  \sin[V(s-t)] 
\,\gamma^{2K}(s-t)
\nonumber \\
&\times&|\gamma(s-t)|^{(1-K)^{2}}\left|
\frac{\gamma^{2}(s+t)}
{\gamma(2t)\gamma(2s)}\right|^{1-K^{2}},
\label{td2}
\eeq
%  \beq
%   A_{0}(t)&=&\int_{0}^{\infty}d\tau \sin
% (Vt)\left[\frac{
% a}{a-iv(t+i\tau)}\right]^{2}  \nonumber \\
% B_{0}(t) &=&i
%  \int_{0}^{t}ds  \sin[V(s-t)] \nonumber \\
%  & \times & \left[
%  \frac{a+iv(s-t)}{a-iv(s-t)}\right]^{K} \left[
%   \frac{a^{2}}{a^{2}+v^{2}(s-t)^{2}}\right]
%   ^{\frac{1+K^{2}}{2}}   \nonumber \\
%   &\times& \left[
% \frac{(a^{2}+4v^{2}t^{2})(a^{2}+4v^{2}s^{2})}
% {[a^{2}+v^{2}(s+t)^{2}]^{2}}\right]^{\frac{1-K^{2}}{4}}
%   \label{td2}
%   \eeq
for $\eta_{I}=0$ and
\beq
A_{1}(t)&=&\sin (Vt)\int_{0}^{\infty}d\tau \,\gamma^{2K}(t+i\tau)
\nonumber \\
B_{1}(t) &=&i
\int_{0}^{t}ds  \,\sin[V(s-t)]\,\gamma^{2K}(s-t)
\label{td1}
\eeq
% \beq
% A_{1}(t)&=&\int_{0}^{\infty}d\tau \sin (Vt)\left[\frac{
% a}{a-iv(t+i\tau)}\right]^{2K} \nonumber \\
% B_{1}(t) &=&i
%  \int_{0}^{t}ds  \sin[V(s-t)]
%  \left[\frac{a}{a-iv(t-s)}\right]^{2K} \label{td1}
%  \eeq
for $\eta_{I}=1$, and where $\gamma(z)=a/(a-ivz)$ and $\xi=\lambda^{2}/(\pi a)^{2}$. 
%Equations (\ref{curregen},\ref{td2},\ref{td1}) and 
%the subsequent discussion constitute the main result of the present work.
In all cases ($\eta_{I},\eta_{T}=0,1$) $j(t)$ is an odd 
function of $V$, as it should be. We also notice that for 
noninteracting systems ($K=1$) we recover the expected result
$A_{1}=A_{0}$ and $B_{1}=B_{0}$. In this case the function 
$\xi \mathrm{Re}[B_{1,0}]$ coincides with the current in 
Eq. (\ref{curr0}) to lowest order in $\lambda$.
We can now provide a quantitative analysis of the TD current 
response for different preparative configurations.

{\em Contacted versus uncontacted ground state.}
\begin{figure}[tbp]
\includegraphics[height=5.cm, width=7.cm]{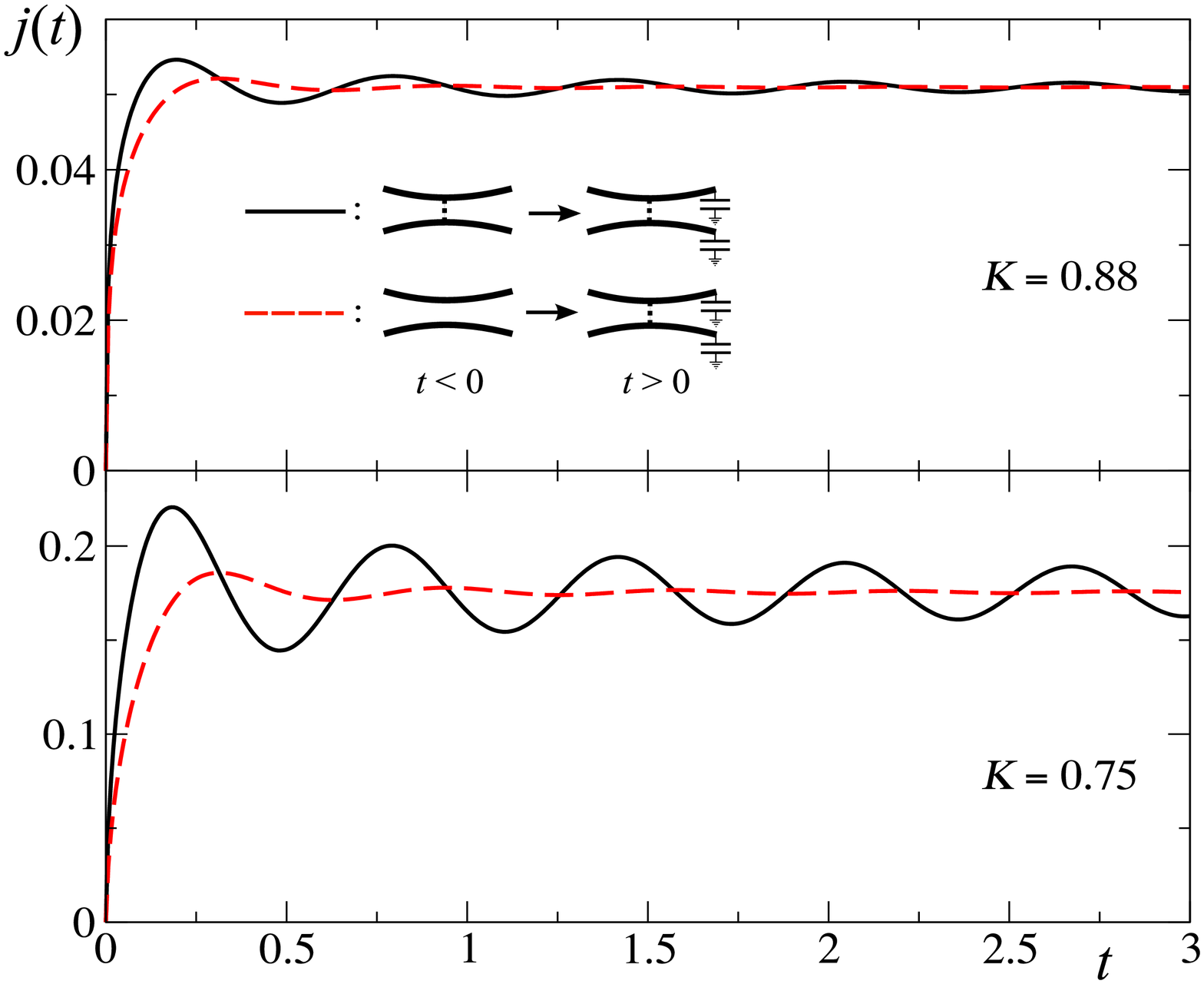}
\caption{Transient currents $j_{T1}(t)$ (solid) and
$j_{T0}(t)$ (dashed) for $V = 10^{-2}$, $K=0.88$ (upper 
panel), and $K=0.75$ (lower panel). 
In the long-time limit they reach the same steady-state value. Current is
in units of $\xi a /v $, $V$ is in units of $v/a$ and $t$ is
in units of $10^{3}a/v$.}
\label{figite}
\end{figure}
We consider an initially contacted ($\eta_{T}=1$) and uncontacted 
($\eta_{T}=0$) correlated ground state ($\eta_{I}=1$) and compare 
the corresponding TD currents
$j_{T1}\equiv \xi \mathrm{Re}[A_{1}+B_{1}]$ and 
$j_{T0}\equiv \xi \mathrm{Re}[B_{1}]$.
The current  $j_{T0}(t)$ has been recently computed in
Ref. \cite{salvay}. In the long time limit it returns the well known
steady-state result 
\be
j_{S}(\b)=\sin(\pi K) \kappa(\beta) \mathrm{sgn}(V)|V|^{\beta}
\label{jsbeta}
\ee
with 
$\kappa(\beta) =- \xi (a/v)^{\beta+1} \Gamma(-\beta) \sin(\beta\pi/2)$
and the exponent $\b=2K-1$, 
obtained long ago by Kane and Fisher\cite{kane}.
Since $A_{1}(t\ra\inf)=0$,  $j_{T1}$ 
approaches the same steady state. 
Note that the small bias limit is ill-defined
for $K<1/2$ due to the break down of the perturbative expansion
in powers of $\l$\cite{feldman,sen}.
Even though $j_{T0}(t\ra\inf)=j_{T1}(t\ra\inf)$ the relaxation is 
 different in the two cases, see Fig. \ref{figite}.
The function  $j_{T0}(t)$ approaches the asymptotic
limit with transient oscillations of frequency $V$ and damping
envelope proportional to $t^{-2K}$ \cite{salvay}. 
The more physical current $j_{T1}$, instead, {\em decays much slower}.
The integral in $A_{1}(t)$  can be
calculated analytically and yields
 \be j_{T1}(t)-
j_{T0}(t)=\xi a^{2K} \frac{\sin (Vt) \cos[(2K-1)\arctan
(vt/a)]}{2v(2K-1)(a^{2}+v^{2}t^{2})^{K-1/2}} , 
\label{trans}
 \ee
which for long times decays as $t^{1-2K}$. (Equation (\ref{trans}) 
provides an independent, TD evidence that the perturbative treatment 
breaks down for $K<1/2$.) Thus, an initially contacted state 
changes the power-law decay from $\sim t^{-2K}$  to
the slower $\sim t^{1-2K}$. 
The amplitude of the
transient oscillations is also significantly different, due to
the factor $(2K-1)^{-1}$ in Eq. (\ref{trans}).  For
$K=0.75$,  $j_{T1}$ oscillates with an amplitude about 10 times 
larger than that of $j_{T0}$, see Fig. \ref{figite}.
The magnification of the oscillations was unexpected since for 
$j_{T1}$ we only switch a bias while for $j_{T0}$ also the contacts.
This effect is not an artifact of the perturbative treatment:
% We verified 
% that this is a genuine 
% effect and not an artifact of the perturbative treatment.
% In fact, 
to support the validity of our results we checked 
that for $\eta_{I}=\eta_{T}=1$ and zero bias the 
density matrix $ \rho(t) =\langle\Q_{0}|
\psi^{\dagger}_{R,H_{1}}(t)\psi_{L,H_{1}}(t)|\Q_{0} \rangle $  
does not evolve in time {\em to first order in $\l$} (this is obvious 
for the exact density matrix). The constant value $\r(t)=\r(0)$
is the result of  a
subtle cancellation of TD functions similar to $A_{1}(t)$ and $B_{1}(t)$.

{\em Correlated versus uncorrelated ground state.} 
\begin{figure}[tbp]
\includegraphics[height=5.cm, width=7.cm]{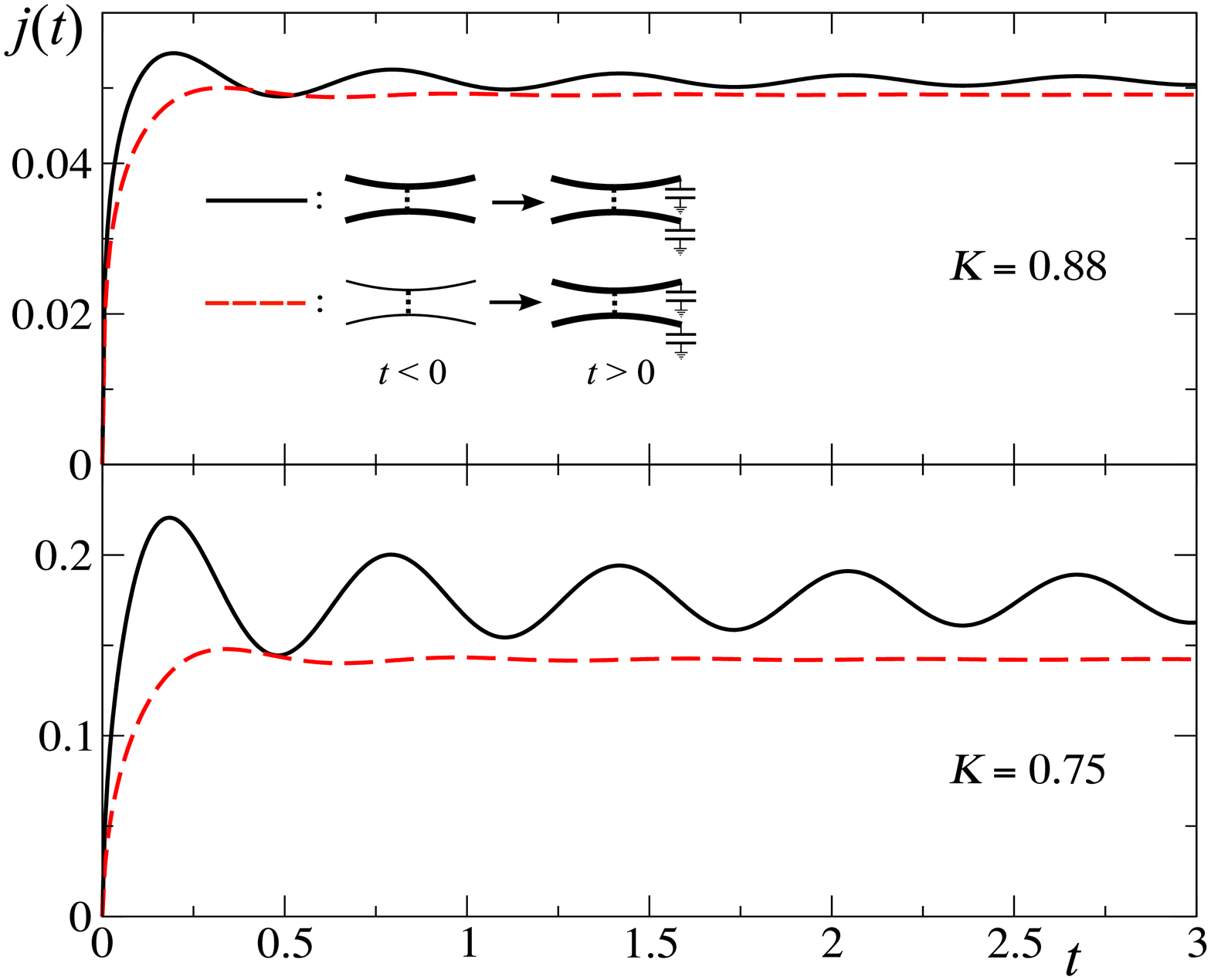}
\caption{Transient currents $j_{I1}(t)$ (solid) and
$j_{I0}(t)$ (dashed) for $V =10^{-2}$, $K=0.75$ (upper panel) 
and $K=0.88$ (lower panel). In the
long-time limit they reach different steady-states. 
Same units as in Fig. (\ref{figite}).}
\label{fig2}
\end{figure}
Next we consider 
the effects of correlations in the ground state. We take $\eta_{T}=1$ 
and compare the TD currents $j_{I1}$ and $j_{I0}$ resulting from Eq. 
(\ref{curregen}) when $\eta_{I}=1$ and $\eta_{I}=0$ respectively. 
Note that $j_{I1}\equiv j_{T1}$ (already calculated
above). The current $j_{I0}= \xi \mathrm{Re}[A_{0}+B_{0}]$ is the response to a sudden bias 
switching and interaction quench; at $t>0$ the electrons
start tunneling from  $L$ to 
$R$ {\em and at the same time} forming interacting quasiparticles.
The interaction quench has a dramatic impact on the
transport properties, both in the transient and steady-state regimes.
From Fig. \ref{fig2} we clearly see that
the relaxation behavior is different. The damping 
envelope of $j_{I0}(t)$ is proportional to $t^{-K^{2}-1}$ as opposed 
to $t^{1-2K}$ of $j_{I1}(t)$. Notice that the exponent $-K^{2}-1<0$ 
for all $K$ (first-order perturbation theory in $\l$ is meaningfull 
for all $K$).

In the long-time limit we find the intriguing result that 
$j_{I0}(t\to\infty)$ is exactly given by
Eq. (\ref{jsbeta}) with exponent $\beta=K^{2}$, thus suggesting that
the structure of the formula (\ref{jsbeta}) is {\em universal}. Below we will prove that this is 
indeed the case and that
$\b$ is an elegant functional of the switching process.
% The nonohmic behavior $\sim V^{K^{2}}$ 
% differs from  $\sim V^{2K-1}$ exhibited by $j_{I1}$.
% The dependence of the steady-state values of $j_{I0}$ and $j_{I1}$ 
% on bias $V$ and interaction strength $K$ is illustrated in 
% Fig. \ref{figcont}.  
For now, we observe that {\em ground state correlations are not reproducible by 
quenching the interaction}.
 The system remembers 
them forever and steady-state quantities are inevitably affected. 
This behavior is reminiscent of the thermalization breakdown
enlightened by Cazalilla\cite{cazalilla} and others\cite{perfetto,man-ekw.2009}.
Here, however, we are neither in equilibrium nor close to it (the  
bias is treated to all orders). The non-equilibrium exponents 
$\b=2K-1$ and $\b=K^{2}$ refer to current-carrying  
states as obtained from the full TD Schr\"odinger equation with 
different initial states.
% Furthermore, our model is not integrable 
% due to the contact perturbation.

\begin{figure}[tbp]
\includegraphics[height=3.5cm, width=7.cm]{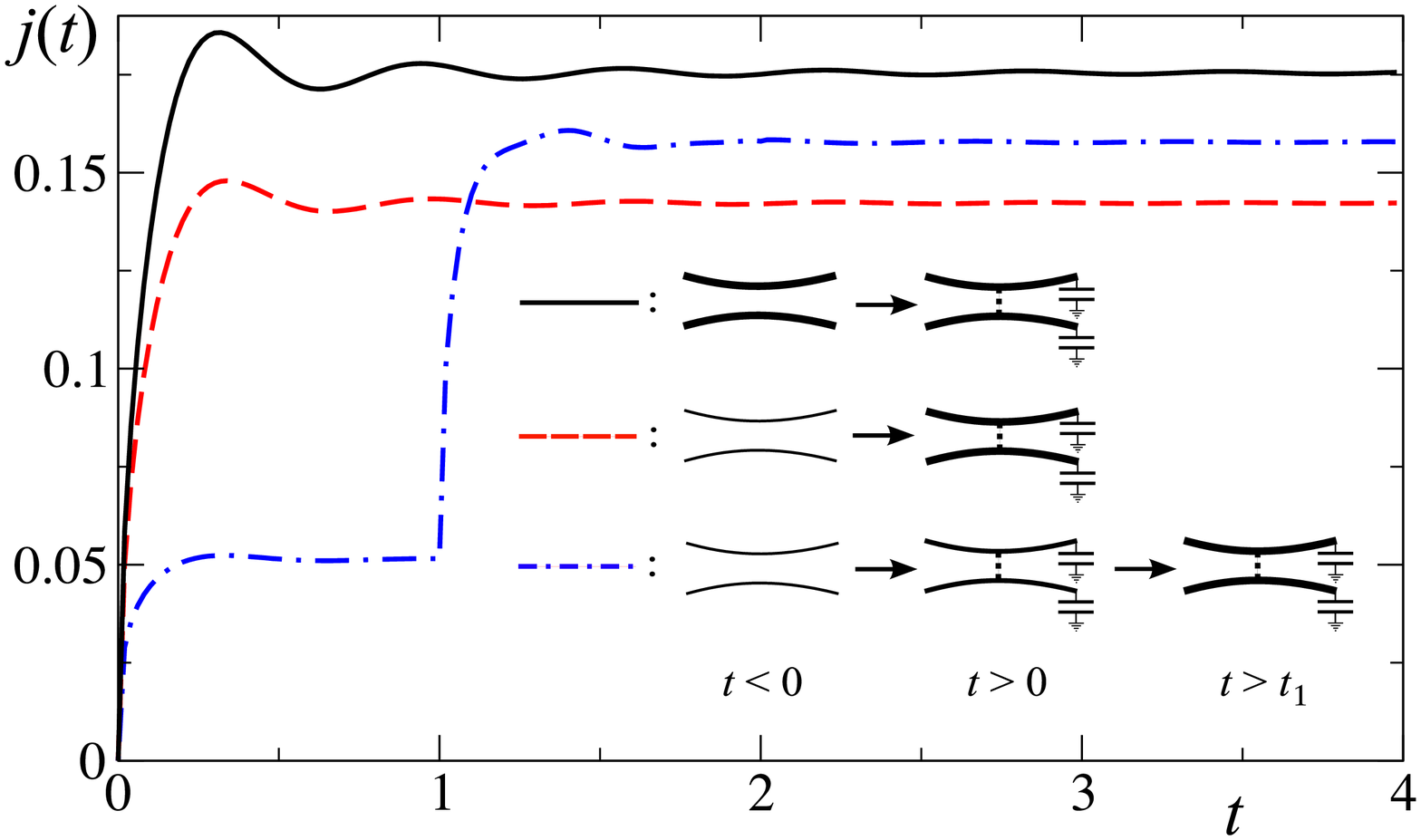}
\caption{Transient currents with $\eta_{I}=1$, $\eta_{T}=0$ (solid), 
and with $\eta_{I}=\eta_{T}=0$ for the quench $1\ra K$
(dashed) and the quench sequence $1\ra \frac{1+K}{2}\ra K$ (dotted-dashed). 
Here $K=0.75$ and $V=10^{-2}$ and the second quench occurs at $t_{1}=1$; 
same units as in Fig. (\ref{figite}).}
\label{figseq}
\end{figure}
{\em History dependence.} 
We now address the question whether or not the physical steady-current 
$j_{S}(2K-1)$ of Eq. (\ref{jsbeta}) is reproducible by more 
sophisticated switching processes of the interaction like, e.g., an 
adiabatic switching. Preliminary insight can be gained by calculating 
$j(t)$ for a double quench: we first quench an interaction with 
$K_{1}=(1+K)/2$, let the system evolve, and then change suddenly
$K_{1}\to K_{2}=K$. The current is calculated along the same 
line of reasoning of Eq. 
(\ref{tdcurr}), although the formulas become 
considerably more cumbersome. In Fig.  
\ref{figseq} we compare the TD currents for initially uncontacted 
leads resulting from an  
interaction $K$ (solid), a single quench $1\to K$ (dashed), and 
the aformentioned double quench (dotted-dashed). 
We clearly see that in the latter case the steady-current is larger 
than $j_{S}(K^{2})$ (single-quench) and gets closer to $j_{S}(2K-1)$.
Strikingly, the double-quench steady-current is again given by $j_{S}(\b)$ of Eq. 
(\ref{jsbeta}) with 
$\b=\frac{1}{2}(1+K_{1}^{2})(1+(\frac{K_{2}}{K_{1}})^{2})-1$.
This value of $\b$ depends only on the $K$-sequence and 
is independent of the quenching times. We have been able to extend 
the above solution to systems initially interacting with $K_{0}$ and 
then subject to an arbitrary sequence of quenches $K_{0}\to 
K_{1}\to \ldots  \to K_{N}=K$. We found the remarkable result that the
formula 
(\ref{jsbeta}) is {\em universal}, with the sequence dependent $\beta$ given 
by
\be
\b[K_{n}]=\frac{K_{0}}{2^{N-1}}\prod_{n=0}^{N-1}
\left[
1+\left(
\frac{K_{n+1}}{K_{n}}
\right)^{2}
\right]-1.
\label{betadis}
\ee
This formula yields the correct values of $\b$ for the single and 
double quench discussed above. Note that for a sequence of increasing 
interactions $K_{n+1}\leq K_{n}$ it holds $\b\geq 2K-1$ with the 
equality valid only for $K_{0}=K_{1}=\ldots=K_{N}=K$. 

We now show that the special value $\b=2K-1$ is  also reproducible by 
an arbitrary (not necessarily adiabatic) {\em continuous} ($N\to 
\infty$) sequential 
quenching. In this limit the variable $x_{n}=n/N$ 
becomes a continuous variable and we can think of the $K_{n}$ as the 
values taken by a differentiable function $K(x)$ in $x=x_{n}$, with $K(0)=K_{0}$ and 
$K(1)=K$. Then, the quantity $\beta$ becomes a functional of $K(x)$ 
that we now work out explicitly. Approximating 
$K(x_{n+1})=K(x_{n}+\frac{1}{N})\approx 
K(x_{n})+\frac{1}{N}K'(x_{n})$ and taking the logarithm of Eq. 
(\ref{betadis}) we can write
\begin{eqnarray}
\log\left(\frac{\b[K(x)]+1}{2K(0)}\right)&=&
\lim_{N\to\infty}\sum_{n=0}^{N-1}\log\left(1+\frac{1}{N}\frac{K'(x_{n})}{K(x_{n})}\right)
\nonumber \\
&=&\int_{0}^{1}dx\frac{K'(x)}{K(x)}=\log\frac{K(1)}{K(0)},
\end{eqnarray}
from which it follows the history independent result
\be
\b[K(x)]=2K-1.
\ee
The above result can easily be generalized to discontinuous switching functions 
$K(x)$ for which, instead, the exponent $\b$ is history dependent.

{\em Conclusions.} In conclusion we studied the role of different preparative configurations in 
TD quantum transport between LLs. 
%First we showed that the currents 
%are analytic functions of the tunneling $\l$ at least for small
%interaction strengths. 
By using bosonization methods
we showed that a sudden switching of the contacts 
does not change the steady-state 
but alters significantly the transient behavior, changing the 
damping envelope from $\sim  t^{-2K}$  to $\sim  t^{1-2K}$ and  
magnifying the amplitude of the oscillations. 
The effects of a sudden interaction quench is even more striking.
Besides a different power law decay ($\sim  t^{1-2K}$ versus $\sim 
t^{-K^{2}-1}$ damping envelope) the steady-current is also 
different; the I-V characteristic $j_{S}\propto V^{\beta}$
changes from $\beta=2K-1$ to $\b=K^{2}$.
%characteristic changes from $j_{S}\propto V^{2K-1}$ to $j_{S}\propto V^{K^{2}}$. 
More generally we proved that for a sequence of interaction quenches 
the steady-current is a universal function of the exponent $\beta$ 
which, in turn, is a functional of the switching process.
It is only for smooth switchings that $\beta$ is history independent 
and equals the value $2K-1$ of the initially interacting LL.
The explicit $\beta$ functional derived in this Letter
establishes the existence of intriguing memory effects that point to
a complex entanglement between equilibrium and non-equilibrium correlations
in strongly confined systems.

%\end{references}


\begin{thebibliography}{99}
\bibitem{note}
Such term stems from first-order perturbation theory along the 
Matsubara time axis. It can be expressed
with mixed GreenÕs functions with one real and one imaginary time
argument, see Refs. \cite{cini-stef,myo}.
\end{thebibliography}
\end{document}